\documentclass[12pt,oneside]{article}

\begin{document}
\thispagestyle{empty}
\setcounter{page}{0}
\renewcommand{\theequation}{\thesection.\arabic{equation}}


\def\a{{\alpha}}
\def\ap{{\a}^{\prime}}
\def\b{{\beta}}
\def\d{{\delta}}
\def\g{{\gamma}}
\def\e{{\epsilon}}
\def\z{{\zeta}}
\def\ve{{\varepsilon}}
\def\vf{{\varphi}}
\def\m\mu
\def\n{{\nu}}
\def\u{{\Upsilon}}
\def\l{{\lambda}}
\def\s{{\sigma}}
\def\t{{\tau}}
\def\th{{\theta}}
\def\vt{{\vartheta}}
\def\nc{noncommutative\ }
\def\npt{non-perturbative\ }
\def\hp{\hat\partial}
\def\kk{{\kappa}}
\def\IR{{\bf R}}
\def\CM{{\cal M}}


{\hfill{LPTENS 02/32}}

{\hfill{NI02009-MTH}}

{\hfill{\tt hep-th/0205115}}

\vspace{2cm}

\begin{center}
{\bf Asymptotic symmetries of ${\rm AdS}_2$ Branes}

\vspace{1.4cm}

Constantin Bachas

\vspace{.2cm}

{\em Laboratoire de Physique Th\'eorique}\\
{\em \'Ecole Normale Sup\'erieure}\\
{\em 24 rue Lhomond, 75231 Paris cedex 05, France} \\
\vspace{.2cm}
{ and} \\
\vspace{.2cm}
{ \em Isaac Newton Institute for Mathematical Sciences
} \\
{ \em 20 Clarkson Road, Cambridge CB3 0EH, UK } \\
\end{center}

\vspace{-.1cm}

\centerline{{\tt bachas@corto.lpt.ens.fr}}

\vspace{1cm}

\centerline{ABSTRACT}

\vspace{- 4 mm}

\begin{quote}\small
I analyze  the asymptotic symmetries of 
a theory of gravity in a 
background consisting of two patches of ${\rm AdS}_3$  spacetime
glued together along an  ${\rm AdS}_2$ brane. 
These  are generated by a  single Virasoro algebra, 
as expected from  the conjectured dual description
in terms of a scale-invariant 
interface separating two conformal field theories. 
Contributed  to the proceedings of
the Francqui Colloquium 2001: `Strings and Gravity:
Tying the Forces Together~.'

\end{quote}

\baselineskip18pt

\newpage


\setcounter{equation}{0}
\section{Introduction}

  A long time ago, Brown and Henneaux \cite{BH} proved that 
a theory of gravity in a
three-dimensional anti-de-Sitter (${\rm AdS}_3$) background has 
an infinite number of asymptotic symmetries, 
generated by two (a left and a right) 
Virasoro algebras   with central charge
\begin{equation}
 c = {3\ell\over 2  G_N}\ .
\end{equation}
\vskip 0.1cm \noindent
Here $\ell$ is the radius of
 ${\rm AdS}_3$,  and $G_N$ the 3D  Newton's constant. 
This is a  remarquable result,  which  shows  how an effect normally 
thought to be  `quantum'  -- the central extension 
of the Virasoro algebras  
--  can arise  from a  classical calculation. 
The result anticipated
the  general ${\rm AdS}_{n+1}$/${\rm CFT}_n$ correspondence \cite{M12}, 
according to which the theory of gravity can be described
by a dual  conformal  theory defined  
at the boundary of  spacetime. It has been rederived by different methods
more recently \cite{HSken,BK,dSS,adam}.

  In this short note I want to  extend the Brown-Henneaux argument to
a situation in which  two different patches of ${\rm AdS}_3$
spacetime  are glued together along a ${\rm AdS}_2$  brane.
One can glue together, more generally, two patches of ${\rm AdS}_{n+1}$
spacetime  along a ${\rm AdS}_n$ brane. 
A related  geometry was studied, for $n=4$,  by
Karch and Randall \cite{KR1}  as a  model  for  localized gravity   
that did not require perfect fine tuning of the brane tension
(see also \cite{kalo,kim,also,Haw}). 
The setup can, furthermore,  be  embedded  in string theory \cite{BP,KR2},
though it is still  unclear whether `realistic models' of  
localized  gravity can be truly obtained in this way.\footnote{See
for instance \cite{B} for a potential obstruction.} 
Quantum gravity in the above background is, in any case,
 believed to admit  a
dual description where  a scale-invariant   interface separates two
(a priori different) conformal theories 
\cite{KR2,BDDO,DFO}.\footnote{Keeping only one ${\rm AdS}_{n+1}$ patch, and treating 
the ${\rm AdS}_n$ as part of the boundary, leads to a
different  holographic picture \cite{KR1,Por}.
I thank Massimo Porrati for pointing this out.}
 For $n=2$ there are no degrees of freedom
on the  interface, which  preserves one (non-chiral)
Virasoro symmetry. What  I will  show here is
how to realize this symmetry in terms 
of non-trivial bulk diffeomorphisms.


\setcounter{equation}{0}
\section{${\rm AdS}_2$  brane in ${\rm AdS}_3$}

   We  consider two patches of ${\rm AdS}_3$  spacetime  glued
together along an ${\rm AdS}_2$  brane.
 The metric in conformally-flat coordinates is
\begin{equation}
ds^2  = {f^{-2}} (dv^2+dy^2-dt^2)\ ,
\label{sol}
\end{equation}
where
\begin{equation}
f(v,y) =
\cases{\  (v + a_1  y)/\ell \;
   & \ \ { for} \ \ $y>0$   \cr \ & \ \cr
\  (v + a_2  y)/\ell \;
   &  \ \ { for} \ \ $y<0$ \ .   \cr}
\label{sol1}
\end{equation}
\vskip 0.25cm \noindent
The coordinates range over all values such that $0< f < +\infty$.
The surface  $f=0$ is the  spacetime boundary, while $f=\infty$ is a
coordinate horizon.
The brane sits at $y=0$ and has radius $\ell$.
The  radii of the bulk ${\rm AdS}_3$  geometry
on either side of the brane are
\begin{equation}
\ell_r = \ell\; {\rm cos}\;\theta_r\ , \ \ \ {\rm where}\ \ \
\theta_r = {\rm arctan}\; a_r\ \ \ {\rm for}\ \ r=1,2~ .
\label{par1}
\end{equation}
One can  check this claim
by rotating  the coordinates   $(v,y)$ by an angle   $\theta_{1(2)}$,
so as to  put  the upper (lower) ${\rm AdS}_3$ 
 patch  in standard Poincar\'e form.
Note that  $\ell_1\not=  \ell_2$  in general,
so that the 3D cosmological constant  jumps discontinuously  at
the position of the brane. This does not happen  if
$a_1=-a_2$ , in which case   $y\to -y$  is a $Z_2$ isometry, or if
$a_1=a_2$ in which case
there is no (back-reacting)  brane. 
The metric and all its derivatives 
are continuous at $y=0$, in the latter  case.

The metric (\ref{sol}--\ref{sol1}) solves the variational equations
that are derived from
the bulk plus brane action
\begin{equation}
S = S^{\rm bulk}_1 + S^{\rm bulk}_2  + S^{\rm boundary}_1 + S^{\rm boundary}_2
+ S^{\rm brane} \ ,
\label{total}
\end{equation}
where
\begin{equation}
S^{\rm bulk}_r  = -{1\over 16\pi G_N}\int_{{\cal M}_r} \sqrt{-g}\; ( R - {2\over
\ell_r^{\ 2}})\
\label{act}
\end{equation}
is the Einstein-Hilbert action with cosmological term(s),
\begin{equation}
S^{\rm boundary}_r  = -{1\over 8\pi G_N}\int_{\partial{\cal M}_r}
\sqrt{-\gamma}\; ( K + {1\over \ell_r} )
\label{boundary}
\end{equation}
is the Gibbons-Hawking boundary term  required
to eliminate second derivatives of the metric \cite{GH},
and
\begin{equation}
S^{\rm brane} =  T\int_{\Sigma}\sqrt{-\hat g}\
\; -\; {1\over 8\pi G_N}\int_{\Sigma} \sqrt{-\hat g}\; [K]  \ .
\label{brane}
\end{equation}
\vskip 0.15cm \noindent
 $\Sigma$ in the above expression is the
worldvolume of the brane,  treated as a thin shell
that  separates spacetime in two disjoint regions
${\cal M}_r$. The boundary of spacetime has been  decomposed as
  $\partial{\cal M}_1 \cup
\partial{\cal M}_2 $ .  The tension of the brane is denoted
$T$,   $\hat g$ is the induced metric on the brane, and
$\gamma$ the metric on the boundary. Furthermore
 $K$  is the trace of the extrinsic curvature tensor,
and $[K]$ its discontinuity across the brane.

The boudary and brane actions  (\ref{boundary})
and (\ref{brane}) require  some
further  explanation. First, the boundaries
$\partial{\cal M}_1 $ and  $\partial{\cal M}_2 $ should be placed
at a  finite cutoff value  $f=\epsilon$, which will  be taken in
the end to zero.
 Following references
\cite{HSken,BK,dSS}, I  have  included in (\ref{boundary})
a counterterm  required to keep  the energy-momentum tensor finite
in this limit.
Secondly, I have
added Gibbons-Hawking surface terms on the
brane wordvolume $\Sigma$,  which is the common boundary of
${\cal M}_1 $ and  ${\cal M}_2 $.
These should be thought as  arising
from the  integral of  the Einstein-Hilbert
action over the (infinitesimal) thickness of the brane.
They do not, therefore,  correspond to a matter source,
even though  I  include
them  for convenience in  the brane action.

Extremizing  the total  action (\ref{total})  leads to
the  Israel junction conditions \cite{Israel,MTW}
(here Latin indices stand for worldvolume directions)
\begin{equation}
[ K_{ab}] - \hat g_{ab} [ K]
\;  = \;   - 8\pi G_N  T \hat g_{ab} \  .
\end{equation}
The extrinsic curvature tensor is the covariant derivative of the
outward-pointing unit normal vector \  ${\hat n}_{\mu} dx^\mu  = - f^{-1} dy$,
projected on the worldvolume of the brane. A straightforward  calculation gives
\begin{equation}
K_{ab}^{(r)} \equiv  - e_a^\mu e_b^\nu\;
\nabla_\mu  {\hat n}_{\nu} = {a_r \over  \ell}\; {\hat g}_{ab}\ ,
\end{equation}
where   $e_a^\mu$ is  a basis
of tangent vectors, and
\begin{equation}
{\hat g}_{ab}\;  dx^a dx^b   = {\ell^2\over v^2}\;(dv^2 - dt^2)\
\end{equation}
is  the induced brane metric. As the reader can verify easily,
the junction conditions are indeed satisfied, provided that
\begin{equation}
 a_1 - a_2 = 8\pi G_N \ell\; T\ .
\label{par2}
\end{equation}
The three free parameters of the metric
(\ref{sol}--\ref{sol1}) can  thus be expressed,
via equations (\ref{par1}) and
(\ref{par2}) ,  in terms of
the  parameters $\ell_r$,  $T$ and $G_N$ that
enter in  the  action.


\setcounter{equation}{0}
\section{Asymptotic symmetries}

   In order to discuss the asymptotic symmetries, it is convenient
to make the following coordinate change~:
\begin{equation}
  u  \equiv  v +  y\;{\rm tan}\theta_r  \;  \ \ {\rm and} \ \
 x \equiv y/{\rm cos}\theta_r  \ ,  
\ \ {\rm in\ \  region}\ \  {\cal M}_r\; .
\label{rep} 
\end{equation}
The new coordinates range
over $0< u < \infty$~,  and $-\infty < x <  \infty$~.
The  brane sits at $x=0$, and  the spacetime boundary is at $u=0$.
As usual, we  introduce an  ultraviolet cutoff in the dual 
conformal theory
by placing the boundary at a finite 
value  $u=\epsilon>0$~.

Although the reparametrization (\ref{rep})  is continuous,  
its Jacobian has a step-function discontinuity at the position of
the brane. Therefore the metric has a corresponding 
step-function jump~:  
\begin{equation}
ds^2 = {\ell^2\over u^2}\; \left( du^2 +dx^2 -dt^2
  -2 \;  {\rm sin}\theta_{r}\; dx\; du \right) \    \ 
\ \ {\rm in\ \  region}\ \  {\cal M}_r\; . 
\label{metr}
\end{equation}
\vskip 0.15cm 
\noindent 
Since this discontinuity is a coordinate artifact,  it is not problematic
so long as one treats it with the required  care.

 The metric (\ref{metr}) has a SL(2,R) group of  isometries, 
which include  time translations, and  the 
global rescalings $(x,t,u)\to \lambda\; (x,t,u)$.  
There is however a larger set of coordinate transformations, which
only leave invariant the asymptotic form of the metric. 
They form a Virasoro algebra, and act on the 
Hilbert  space of states  at the boundary.  
The relevant infinitesimal transformations (in region ${\cal M}_r$) are:
\begin{equation}
 x^\pm \rightarrow x^\pm - \xi^\pm +
 { u\over 2}\; {\rm sin}\theta_r\;
 ( \partial_+\xi^+ - \partial_-\xi^-)
- { u^2\over 2} \; \partial_{\mp}^2 \xi^{\mp}\ , 
\label{j1}
\end{equation}
and
\begin{equation}
 {1\over u}  \rightarrow {1\over u} + {1\over 2u}\;
(\partial_+\xi^+ + \partial_-\xi^-) - { {\rm sin}\theta_r\over 2}\;
(\partial_+^2\xi^+ -\partial_-^2\xi^-)\ .
\label{j2}
\end{equation}
\vskip 0.15cm \noindent  Here $x^\pm = t\pm x$
are light-cone coordinates, and $\xi^\pm = f(x^\pm)$ 
with $f$ an arbitrary  infinitesimal  function. 
The reader will have no problems  verifying that the above  transformations~:
(a)  act on the boundary, at $u=0$,  as  conformal mappings
which  preserve  the $x=0$ interface, 
and (b) that they reduce 
to the Brown--Henneaux transformations \cite{BH} 
in the case of  pure anti-de-Sitter spacetime,
 i.e. for $\theta_r=0$~.

  To check  that they are  asymptotic symmetries, we need
to calculate the transformed metric. A lengthy but straightforward
computation gives the following variation in region ${\cal M}_r$~:
\begin{eqnarray}
2\; ds^2 \rightarrow & 2\; ds^2 + \ell^2 {\rm cos}^2\theta_r \; [
\partial_-^3\xi^- (dx^-)^2 + \partial_+^3\xi^+ (dx^+)^2] \nonumber \\
\label{metrictr}
&+ \; \ell^2 {\rm sin}^2\theta_r \; (\partial_+^3\xi^+ +
\partial_-^3\xi^-)\; dx^+ dx^-   \\
 &+\; \ell^2  {\rm sin}\theta_r\;
(\partial_+^3\xi^+ dx^+ - \partial_-^3\xi^- dx^-)\; du\
\nonumber .
\end{eqnarray}
\vskip 0.15cm \noindent 
We see that the metric variation is down by two powers of $u$. 
The transformations (\ref{j1}--\ref{j2}) correspond therefore
to asymptotic symmetries, if we supplement the definition
of the theory with  the boundary conditions  $\delta g_{\mu\nu}
\sim o(1)$ near  $u\to 0$.\footnote{Our boundary conditions look
superficially  weaker than the ones 
of Brown and Henneaux, who
require  the off-diagonal components $g_{u\;\pm }$  to vanish linearly.
This can,   however, be always  arranged,  
so long as ${\rm sin}\theta_r \not= 0$, 
with the help of additional (subleading) coordinate transformations: 
$\delta (1/u) \sim o(u)$ and $\delta t \sim o(u^3)$.}

 It is worth stressing that 
$\xi^+(t) =  \xi^-(t)$ is required by continuity of  the
coordinate transformations 
at the location of the brane,  at $x=0$. The same condition also forces
 the  infinitesimal function $f$ 
to be the same  in the regions ${\cal M}_1$ and ${\cal M}_2$.
The asymptotic symmetries therefore  depend on a single
 arbitrary function. 
This agrees with  the holographic interpretation,  
in which only a single   Virasoro symmetry survives \cite{BDDO}.


\setcounter{equation}{0}
\section{Energy-momentum tensor}

To further elucidate the meaning of the asymptotic  symmetries,
we will compute their effect  on  the 
energy-momentum tensor. Following \cite{BY, BK} we define this latter
 as the variation of the total action with respect to
the  metric at the boundary. For solutions of the classical equations
the bulk terms  in the action don't  contribute,
so we  find  
\begin{equation}
T_{ab}\;  =\;  {2\over \sqrt{-\gamma}}
 {\delta S\over \delta\gamma^{ab}}\;  =\;   {1\over 8\pi G_N}
\left[ K_{ab} - K \gamma_{ab} - {1\over \ell_{r}}\gamma_{ab}\right]  \ .
\label{emtensor}
\end{equation}
Latin indices here refer to the two boundary coordinates $x$ and $t$, and
they are  raised and lowered with $\gamma$. They should not be confused with
the  brane-worldvolume indices of section 2.

 It is convenient to write  the 3D metric in  ADM
form
\begin{equation}
ds^2 = \gamma_{ab}(dx^a + N^a du)(dx^b +N^b du) + (N du)^2\ .
\end{equation}
The extrinsic curvature can then be expressed in terms of the
lapse and shift functions as follows \cite{MTW}~:
\begin{equation}
K_{ab} = - {1\over 2N} \left[ \partial_{a} N_{b}
+ \partial_{b} N_{a} - 2 N^c \Gamma_{c\vert ab}
- {\partial\gamma_{ab}\over \partial u}  \right] \ ,
\label{Kadm}
\end{equation}
with
\begin{equation}
\Gamma_{c\vert ab} = {1\over 2} ( \partial_a\gamma_{bc} +
\partial_b\gamma_{ac} - \partial_c\gamma_{ab})\ .
\end{equation}
For the metric (\ref{metr}) we have
\begin{equation}
\gamma_{ab} = {\ell^2 \over u^2}\;  \eta_{ab}\ , \ \ \
N^x = - {\rm sin}\theta_{r}\ , \ \ \ {\rm and}\ \ \
N = {\ell_{r} \over u}\ .
\end{equation}
A simple calculation then gives
\begin{equation}
K_{ab} = -{1\over \ell_{r}}\; \gamma_{ab}\ \ \Longrightarrow \ \
T_{ab} = 0\ .
\end{equation}
This is in acoordance with the fact that the vacuum expectation
value of the energy-momentum tensor in the dual CFT should vanish.

To calculate the transformed energy-momentum tensor, we first put
(\ref{metrictr}) in 
 ADM form. The variation   $\delta\gamma_{ab}$ can be 
read off directly
from  the transformed metric. The variation of the lapse and
shift functions can be derived easily with the result:
\begin{equation}
\delta N^\pm = \pm {u^2}\;{\rm sin}\theta_r\;
\left[{3\over 2}{\rm cos}^2\theta_r \; \partial_\mp^3\xi^\mp -
{1\over 2}{\rm sin}^2\theta_r \; \partial_\pm^3\xi^\pm\right]\ , 
\end{equation}
and
\begin{equation}
\delta N = {u \over 2}\;\ell_r\; {\rm sin}^2\theta_r
\; (\partial_+^3\xi^+ +
\partial_-^3\xi^-)\ .
\end{equation}
From  expression (\ref{Kadm}) 
and the fact that $\delta\gamma_{ab}$ is independent
of $u$,   we find the following variation of  the extrinsic curvature:
\begin{equation}
\delta K_{ab} = -{\delta N\over N} K_{ab} -{1\over 2N} \left(
\gamma_{ac}\;\partial_b\delta N^c+ \gamma_{bc}\;\partial_a\delta
 N^c +N^c\partial_c\delta\gamma_{ab}\right)
\ .
\end{equation}
Inserting the expressions for $\delta\gamma_{ab}$, $\delta N^\pm$
and $\delta N$  leads to:
\begin{equation}
\delta K_{+-} = -{\ell^2 
{\rm sin}^2\theta_r\over 4\ell_r}\; (\partial_+^3\xi^+ +
\partial_-^3\xi^-)\ , \ \ \ 
\delta K = 0\ ,
\end{equation}
and 
\begin{equation}
\delta K_{\pm\pm} = \mp {u \ell^2 \over 2\ell_r}\;{\rm sin}\theta_r\;
{\rm cos}^2\theta_r\; \partial^4_\pm\xi^\pm\ . 
\end{equation}
Furthermore from equation (\ref{emtensor}) we have~: 
\begin{equation}
8\pi G_N \; \delta T_{ab} = \delta K_{ab} + 
{1\over \ell_r} \delta\gamma_{ab}\ .
\end{equation}
Substituting our  results for $\delta K_{ab}$ and $\delta\gamma_{ab}$,
and dropping terms that vanish at $u=0$,
leads to  our  final expression valid in the region ${\cal M}_r$:
\begin{equation}
\delta T_{\pm\pm} =  {\ell_r \over 16\pi G_N}\;
 \partial_\pm^3 \xi^\pm \ , \ \ \ {\rm and} \ \ \ \
\delta T_{+-} = 0\ . 
\label{final}
\end{equation}

  Recall  the usual transformation of
the energy-momentum tensor under a  2d conformal map~:
\begin{equation}
\delta T_{\pm\pm} = - (2\partial_\pm \xi^\pm  +
 \xi^\pm\partial_\pm) T_{\pm\pm} + {c\over 24\pi} \partial^3_\pm \xi^\pm\ . 
\end{equation}
Comparing with (\ref{final}) we conclude that the 
central charge, 
\begin{equation}
c_r = {3\ell_r\over 2G_N}\ \ \ {\rm in}\ \ {\cal M}_r\ ,
\end{equation}
must jump  discontinuously at $x=0$. This is again
consistent with the dual description,  where two (a priori) distinct
conformal theories are glued together along a common interface.

\vspace{9mm}

\noindent {\bf Acknowledgments}:
I thank the organizers of the Francqui colloquium, and in
particular Marc Henneaux and Alex Sevrin, for the invitation
to a truly memorable  meeting. 
My warmest  congratulations  to Marc
for a well-deserved prize, along with  best wishes for the future.

\end{document}